\documentclass[preprint,showpacs,amsmath,amssymb, letterpaper, superscriptaddress, prb]{revtex4}
\usepackage{graphicx}
\usepackage{dcolumn}
\usepackage{subfigure}
\usepackage{bm}

\newcommand{\rt}{{\sqrt{3}}}

\begin{document}

\title{Novel Ground-State Crystals with Controlled Vacancy Concentrations: From Kagom\'{e} to Honeycomb to Stripes }
\author{Robert D. Batten}
\affiliation{Department of Chemical Engineering, Princeton University, Princeton, NJ 08544, USA}

\author{David A. Huse}
\affiliation{Department of Physics, Princeton University, Princeton, NJ, 08544, USA}

\author{Frank H. Stillinger} 
\affiliation{Department of Chemistry, Princeton University, Princeton, NJ, 08544, USA}

\author{Salvatore Torquato}
\altaffiliation{Corresponding author}
\email{torquato@electron.princeton.edu} 
\affiliation{Department of Physics, Princeton University, Princeton, NJ, 08544, USA}
\affiliation{Department of Chemistry, Princeton University, Princeton, NJ, 08544, USA}
\affiliation{Princeton Institute for the Science and Technology of Materials, Princeton University, Princeton, NJ, 08540, USA}
\affiliation{Program in Applied and Computational Mathematics, Princeton University, Princeton, NJ, 08544, USA} 
\affiliation{Princeton Center for Theoretical Science, Princeton University, Princeton, NJ, 08644, USA}
\affiliation{School of Natural Sciences, Institute for Advanced Study, Princeton, NJ, 08544 USA}

\date{\today}

\begin{abstract}
We introduce a one-parameter family, $0 \leq H \leq 1$, of pair potential functions with a single relative energy minimum that stabilize a range of vacancy-riddled crystals as ground states.  The ``quintic potential'' is a short-ranged, nonnegative pair potential with a single local minimum of height $H$ at unit distance and vanishes cubically at a distance of $\rt$.  We have developed this potential to produce ground states with the symmetry of the triangular lattice while favoring the presence of vacancies.  After an exhaustive search using various optimization and simulation methods, we believe that we have determined the ground states for all pressures, densities, and $0 \leq H \leq 1$.  For specific areas below $3\rt/2$, the ground states of the ``quintic potential'' include high-density and low-density triangular lattices, kagom\'{e} and honeycomb crystals, and stripes. We find that these ground states are mechanically stable but are difficult to self-assemble in computer simulations without defects.  For specific areas above $3\rt/2$, these systems have a ground-state phase diagram that corresponds to hard disks with radius $\rt$.  For the special case of $H=0$, a broad range of ground states is available. Analysis of this case suggests that among many ground states, a high-density triangular lattice, low-density triangular lattice, and striped phases have the highest entropy for certain densities.  The simplicity of this potential makes it an attractive candidate for experimental realization with application to the development of novel colloidal crystals or photonic materials.
\end{abstract}

\maketitle

\section{Introduction}

The field of self-assembly provides numerous examples of using atoms, molecules, colloids, and polymers as building blocks in the development of self-organizing functional materials. For example, researchers have used nanoparticles to create stacked rings and spirals using DNA linkers,\cite{sharma2009control} to construct photonic crystals using binary systems of colloids,\cite{hynninen2007self} and to prototype a self-organzing colloidal battery.\cite{cho2007self}

Recently, there have been significant efforts to design pair potentials that yield self-assembly of targeted ground-state (potential energy minimizing) structures using ``inverse methods.''\cite{torquato2009inverse, cohn2009algorithmic}  With an inverse approach, one chooses a targeted structure or property and uses optimization methods to develop a robust pair potential function. We envision that colloids will be the experimental testbed for the optimized interactions because colloids offer a significant range of repulsive and attractive interactions that can be tailored in the laboratory.\cite{russel1989colloidal}  For these models, the many-body interactions of a system are reduced to pair interactions.  For a pair potential $v(r)$, where $r$ is the distance between two particles, the potential energy per particle $u$ of a many-body configuration ${\bf r}^N$ of $N$ particles reduces to $u({\bf r}^N) = \frac{1}{N}\sum_{i<j}v(r_{ij})$.

Inverse approaches have been applied to produce low-coordinated ground states such as honeycomb,\cite{rechtsman2005optimized} square,\cite{rechtsman2006designed} simple cubic,\cite{rechtsman2006self} diamond and wurtzite crystals\cite{rechtsman2007synthetic} as well as many-body configurations on a sphere.\cite{cohn2009algorithmic} In addition to targeting material structures, the inverse approach has been applied to material properties including negative Poisson's ratio,\cite{rechtsman2008negative} negative thermal expansion,\cite{rechtsman2007negative} and scattering characteristics of many-particle systems.\cite{batten2008classical}  The ability to control vacancy concentrations and arrangements in ground-state structures is one property yet to be achieved via inverse methods. We expect the design of vacancy-riddled crystals to have important technological applications and may provide fundamental insight into certain physical phenomena. The scattering of light,\cite{joannopoulos2008photonic} ionic conductivity,\cite{agrawal1999superionic} transport processes in heterogeneous materials,\cite{torquato2002rhm} and possibly supersolid behavior in quantum systems\cite{andreev1969soviet,kim2004probable} are directly related to vacancies in a material structure.

\begin{figure}
\includegraphics[width=0.6\textwidth, clip=true]{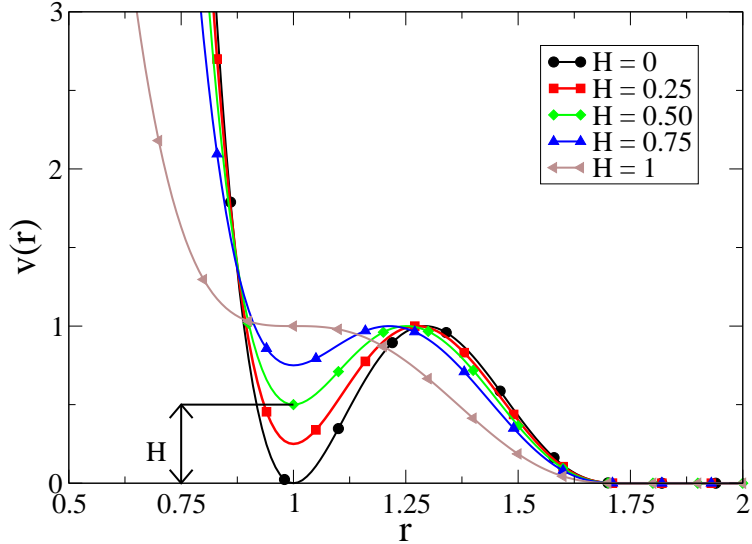}
\caption{(Color online) Quintic potential, Eq.\ \ref{eq:potential} for various $H$ values.  The scales of energy and length are dimensionless.}
\label{fig:potential} 
\end{figure}

In this paper, we introduce the ``quintic potential'' as a pair interaction function that yields vacancy-riddled lattices and low-coordinated crystals as ground states at high density and striped patterns as ground states at low-density. Although not a result of a true inverse method, this family of potential functions arose through a selection process so that vacancy-riddled and low-coordinated lattices are energetically degenerate or possibly favorable to the triangular lattice for certain densities. Each quintic potential function curve, illustrated in Fig.\ \ref{fig:potential}, is steeply repulsive for small $r$, has a local minimum at $r=1$ so that $v(1)=H$, is nonnegative, and is smoothly truncated to be zero at $r=\rt$ and beyond.  The algebraic form of the potential is given in Sec.\ \ref{sec:potential}.  The understanding of the phase behavior associated with the quintic potential is the first step toward developing a more comprehensive inverse approach to controlling vacancies in ground-state structures.  

We find that ground states can vary from a high-density triangular lattice, the kagom\'{e} and honeycomb crystals, a striped phase, a low-density triangular lattice, and a hard-disk-like fluid as the pressure decreases.  The positivity of the local relative minimum is a key feature that allows vacancies to be stabilized in the ground state. Consider a system of particles arranged in a triangular lattice at a specific area (area per particle) $a =\rt/2$ so that each of the particle's first neighbors lie at $r=1$ and second neighbors at $r=\rt$.  Because the local minimum in $v(r)$ is at positive energy, the removal of a particle will lower the total potential energy of the system.  However, to maintain the constraint on $a$, the lattice spacing must decrease, which increases the total potential energy.  This family of potentials is designed so that the creation of a vacancy and the subsequent reduction in lattice spacing yields a net decrease in the potential energy per particle.  

However, the general control of vacancies is a challenging task due to long-ranged vacancy-vacancy interactions. For the Lennard-Jones and many short-ranged repulsive potentials, vacancies arise in equilibrium solids at positive temperature due to their entropic contribution to the free energy.\cite{ashcroft1976ssp} At $T=0$, they can be ``frozen in'' during cooling, though they are not present in the equilibrium ground states of these systems.  For typical potentials whose ground states are ordered, some pair potentials, such as the Lennard-Jones interaction, vacancies cost energy due to missing ``bonds'' of negative energy, while with repulsive potentials and the electron crystal, vacancies cost energy due to strain energy in the crystal. 

If a vacancy is introduced into an otherwise perfect crystal at positive pressure, strains will arise so that the system can reduce its potential energy and achieve mechanical stability.  In typical materials such as Lennard-Jones systems, these relaxations cannot reduce the energy below that of the ground state. When more than one vacancy is present, the strain fields interact and subsequently mediate long-ranged vacancy-vacancy interactions.  These interactions are highly nontrivial even in the case of two vacancies in a crystal. In two dimensions, several studies have examined these effective interaction energies between two vacancies as a function of separation distance for a number of potentials including the electron crystal (Coulombic potential with positive background charge),\cite{fisher1979defects, cockayne1991energetics,candido2001single} Lennard-Jones,\cite{modesto2008interaction} Gaussian core,\cite{lechner2009point} and a repulsive $r^{-3}$ potential.\cite{lechner2009defect}

In the two-dimensional electron crystal, vacancy-vacancy interactions are attractive and fall off as $r_v^{-3}$,\cite{cockayne1991energetics} where $r_v$ is the separation distance between two vacancies, which agrees with continuum elasticity theory.\cite{fisher1979defects} For the Lennard-Jones potential, vacancy-vacancy interactions are also attractive at least for short distances.\cite{modesto2008interaction} These effective vacancy interaction potentials are not necessarily monotonic as a function of lattice spacing.\cite{candido2001single, modesto2008interaction} The displacement fields can be highly anisotropic near a vacancy.\cite{lechner2008displacement,lechner2009point} In the vicinity of the vacancy, the atomistic details account for the discrepancy between numerical results for the displacement fields and those predicted by continuum elasticity theory.  Beyond fifteen lattice spacings, this continuum theory accurately reproduced the results of numerical studies under some boundary conditions.\cite{lechner2009point}

The arrangement of low concentrations of vacancies may not simply be controllable because of the nontrivial coupling between the pair potential function, the local deformations that it produces when a vacancy is introduced into a perfect crystal, and the effective vacancy-vacancy interactions.  However, a number of pair potential functions have been found to give rise to ground states with high concentrations of vacancies. For example, the ``honeycomb potential'' was developed via inverse methods to yield the honeycomb crystal as a robust ground-state structure.\cite{rechtsman2005optimized, rechtsman2006designed} At positive temperatures, the honeycomb structure remained stable for a large temperature and pressure region of the phase diagram.\cite{hynninen2006global} This double-well potential was generalized as the Lennard-Jones-Gaussian potential, which showed a number of complex crystal structures including pentagonal, hexagonal, nonagonal, and decagonal unit cells as the depths and locations of the energy wells were varied.\cite{engel2007self,engel2009structural} Colloidal particles with three ``patches,'' or attractive interaction sites on the surface of a particle, have recently been shown to yield a honeycomb structure for some pressures at $T=0$ in addition to other low-coordinated crystal structures.\cite{doppelbauer2010self}  In addition, the square-shoulder potential, a hard-core potential with a soft corona of variable length, shares some of the same ground-state features as the quintic potential.  In particular, low-coordinated structures such as a honeycomb-like trivalent configuration and lanes are ground-state features\cite{fornleitner2008lane, fornleitner2010pattern} that are shared with the quintic potential.  Such a potential can also form lamellar and micellar phases for relatively long-ranged coronas as was shown theoretically.\cite{glaser2007soft}

Lastly, we note that several potentials qualitatively similar to the quintic potential have been studied in other contexts. For example, a potential that gives rise to quasiperiodic order and the square lattice in two dimensions has one local minimum and local maximum and finite cutoff,\cite{quandt1999formation} as does the Dzugutov potential\cite{dzugutov1992glass} which gives rise to a number of complex local clusters in three dimensions.\cite{doye2001global}  However, both of these potentials have a local attractive minimum ({\it i.e.} negative potential energy).  A piecewise linear potential with a hard core, single local minimum, and single local maximum showed the formation of chains and labyrinths at positive temperature, though the ground state is not fully characterized.\cite{haw2010growth} However, in contrast to the above potentials, the quintic potential is short-ranged, positive, isotropic, continuous and differentiable, which may be beneficial to experimentalists hoping to realize this potential in a laboratory setting.  We find that with this potential, a number of low-coordinated structures arise as ground states including the honeycomb and kagom\'{e} crystals as well as a ``striped'' phase.  This represents a significant achievement and a first step toward controlling vacancies in ground-state structures.

The remainder of this paper is as follows.  The quintic potential is defined and detailed in Sec.\ \ref{sec:potential} while our methodology is detailed in Sec.\ \ref{sec:methods}.  In Sec.\ \ref{sec:phases}, we define the phase diagram in the specific area-$H$ plane and in the pressure-$H$ plane. We discuss the characteristics of the phases, the results of simulation, and their mechanical stability.  In Sec.\ \ref{sec:cases}, we focus on the special case where $H=0$. This case yields anomalous behavior because the potential energy vanishes inside and outside the energy barrier.  Lastly, we discuss the implications of this potential and identify extensions to this work in Sec.\ \ref{sec:discussion}.

\section{Quintic Potential}
\label{sec:potential}

\begin{figure}
\includegraphics[width=0.4\textwidth, clip=true]{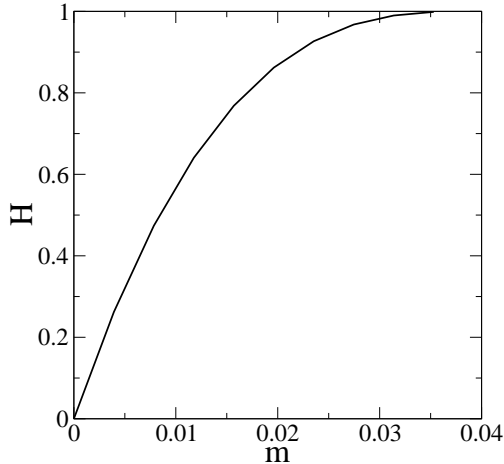}
\caption{ The relationship between $m$ and $H$ for the quintic potential.}
\label{fig:mvh} 
\end{figure}

The generalized quintic potential consists of a fifth-order polynomial that is truncated beyond $\rt$ to be zero. It has the form
\begin{equation}
f(r) = \left[ L(m) \left(r-\sqrt{3}\right)^5 + K(m)\left(r-\sqrt{3}\right)^4- \left(r-\sqrt{3}\right)^3\right], \quad r\leq\rt
\end{equation}
and zero otherwise, where $m$ is an unscaled height of the local minimum.  The coefficients $L(m)$ and $K(m)$ are chosen so that $r=1$ is a local minimum and $f(1) = m$, and, as a function of $m$, are given respectively as
\begin{eqnarray}
 L(m) &=& 4m\left(\sqrt{3}-1\right)^{-5} - \left(\sqrt{3}-1\right)^{-2},\\
 K(m) &=& 5m\left(\sqrt{3}-1\right)^{-4}-2\left(\sqrt{3}-1\right)^{-1}. 
\end{eqnarray}
The condition that $f''(1) >0 $ ensures that the stationary point is a relative minimum, and therefore the parameter $m$ must be constrainted to 
\begin{equation}
 m < \frac{\left(\sqrt{3} - 1\right)^3}{10} \approx 0.0392304845.
\end{equation}
The position of the energy barrier (the relative maximum) occurs at
\begin{equation}
 r_{b} = \sqrt{3} - \frac{3\left(\sqrt{3} - 1\right)}{5\left[1-4m\left(\sqrt{3}-1\right)^{-3}\right]}.
\end{equation}
The generalized pair potential is rescaled so that $v(r_{b})= 1$ to set a uniform energy scale,
\begin{equation}
 v(r) = f(r)/f(r_{b}).
\label{eq:potential}
\end{equation}
Upon rescaling, the height of the relative minimum $H$ is defined so that $v(1)=H$ and varies from zero to unity.  When $H=1$, the potential is monotonically decreasing and is flat at $r=1$. In this case, there is no local minimum or maximum. These soft potentials are repulsive for small $r$, and the first and second derivatives vanish at $r=\rt$.   We construct the potential such that the first and second derivatives vanish at $r=\rt$ so that the second-order expansions of the potential energy required for stability calculations ({\it e.g.} phonons) are well-defined. Examples from the family of quintic potentials are shown in Fig.\ \ref{fig:potential}.  Because developing a simple expression relating $m$ and $H$ is difficult, a table relating $m$ and $H$ was used. The relation between $m$ and $H$ is shown in Fig.\ \ref{fig:mvh}.

\section{Methods}
\label{sec:methods}

At $T=0$, the free energy per particle $g$ is identical to the enthalpy per particle and is related to the average potential energy per particle $u$ via $g=u+pa$. To map the ground-state phase diagram as a function of specific area $a$, pressure $p$, and $H$, we utilized the double-tangent construction method to find the lowest free-energy structure among candidate ground-state configurations. The double-tangent construction is demonstrated in Fig.\ \ref{fig:latticesums-h0.5}.  

We considered several crystal structures as candidate ground states including the triangular lattice (TRI), square lattice (SQ), honeycomb crystal (HC), kagom\'{e} crystal (KAG), and its close relative, the ``anti-kagom\'{e}'' crystal (AKG).  Whereas the kagom\'{e} crystal has vacancies located on a triangular lattice separated by two nearest neighbor spacings, the anti-kagom\'{e} crystal has vacancies located on a rectangular lattice.  For all densities, the anti-kagom\'{e} has a potential energy greater than or equal to the kagom\'{e} lattice due to the differing local coordination numbers.  We considered these crystal structures because they are relative simple crystal structures with varying degrees of local coordination. The construction of the potential, with a well at $r=1$ and vanishing energy at $r=\rt$ would intuitively favor such geometries. We use other techniques to systematically explore other crystals with a small number of particles in the unit cell. With these crystal structures, we performed lattice sums over the relevant range of specific area.

We also performed an exhaustive search for energy-minimizing $n$-particle crystals, periodic configurations containing $n$ particles per unit cell, while allowing for shape deformation of the unit cell. For example, consider a two-particle crystal with lattice vectors $[d_1, 0]$ and $[d_2, d_3]$ and a basis where particle ${\bf 0}$ is at $(0, 0)$ and particle ${\bf 1}$ is located at $(x_1, y_1)$.  For this system, the potential energy per particle is given as 
\begin{equation}
 u = v({\bf r_{01}}) + \frac{1}{2}\sum_{\bf i} \left[v({\bf r_{00'}}) + v({\bf r_{11'}}) + v({\bf r_{01'}})+ v({\bf r_{10'}})\right],
\end{equation}
where the summation is over all lattice sites except the origin site and ${\bf 0}'$ represents particle ${\bf 0}$ in a cell other than the origin cell (and likewise for particle ${\bf 1}$).  Minimizing $u$ for a fixed area per particle $a$ and eliminating $d_3 = na/d_1$ casts this as an unconstrained minimization that is function of $d_1$, $d_2$, $x_1$ and $y_1$.  The function is minimized using the conjugate gradient algorithm.  It is straightforward to generalize this to a larger number of particles in the unit cell.

To ensure that we obtain globally optimal solutions, we use a large number of initial conditions slowly and systematically varying $d_1$, $d_2$, $x_1$ and $y_1$ from zero to $n\rt$.  After each minimization, we ensure that the unit cell is not significantly sheared. A highly sheared unit cell requires summation over a large number of unit cells to ensure that all interactions are accounted for and therefore we discard those solutions with angles less than 10$^\circ$. This procedure is repeated for nearly two thousand values of $a$ on the range $0.7 \leq a \leq 3\rt/2$ so that a smooth $u$ versus $a$ curve is generated.  

These minimizations are a function of $2n$ variables, and therefore many minimizations at a fixed area per particle can be performed easily. However, as the number of particles per unit cell grows, the number of initial conditions required to ensure global optimality grows as $2^{n}$. We performed these minimizations systematically for up to four particles per unit cell, beyond which it becomes too intense to systematically explore thousands of initial conditions for thousands of specific areas.

We also used slow cooling via molecular dynamics (MD) to obtain candidate ground-state structures for larger and possibly disordered systems. With a system of 800 particles in a periodic box, the system was simulated using the Verlet algorithm and Andersen thermostat.\cite{frenkel2001understanding} The system was initialized as a high-temperature liquid and slowly cooled according to a linear temperature schedule from $T=0.4$ to 0.025 over the course of 1.6$\times10^7$ time steps.  After the simulation was terminated, the system was quenched to a potential energy minimum using the conjugate gradient algorithm. 

Lastly, we performed lattice Monte Carlo (MC) optimizations using several hundred lattice sites. In these optimizations, the specific area was fixed and the lattice sites were either occupied or unoccupied by particles. The optimization variables were the occupation state of the lattice sites, the angle between the lattice vectors and the length of one lattice vector. The length of the other lattice vector was constrained by the area, number of particles, angle between the lattice vectors, and the length of the first lattice vector. Possible Monte Carlo moves included inserting a particle to a vacant site, removing a particle from an occupied site, swapping a particle from an occupied site to an unoccupied site, perturbing the angle of the lattice, and perturbing the length of one of the lattice vectors. The system was then optimized via simulated annealing. It was given an initial ``temperature,'' or energy scale, and moves were accepted or reject based upon the Metropolis algorithm. Ten thousand MC trials were attempted in each cycle and several hundred cycles were performed for each optimization.

It is important to note that although many interesting structural characteristics arose from the results of molecular dynamics simulation and Monte Carlo optimization, no final configurations were identified as ground-state structures. The lattice sums and crystal optimization methods always provided the lowest free-energy structures.

The double-tangent construction requires care and precision since other phases can come very close to the coexistence free energy.  The $p$-$g$ and $u$-$a$ curves are discretized over several thousand data points and linearized between between data points. The identification of coexisting phases can be done by using the lower, concave envelope of the $p$-$g$ curve. The coexistence points and ranges of stability were then calculated using a tabulation of $u$, $a$, $p$ and $g$ and linear interpolation.

\section{Results}
\label{sec:phases}

\subsection{Phase Diagram}

\begin{figure}
\includegraphics[width=0.6\textwidth, clip=true]{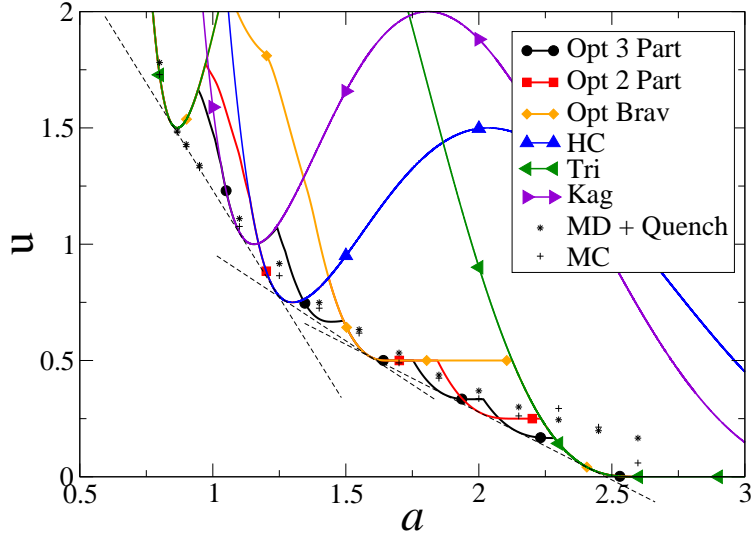}
\caption{(Color online) Potential energy per particle $u$ as a function of specific area $a$ for selected crystals and the optimal 2- and 3-particle crystals for $H=0.5$.  The dense triangular, honeycomb, striped, and open triangular crystals are stable ground states for $H=0.5$. The square and anti-kagom\'{e} structures are omitted for clarity. Dashed lines represent the double tangent construction.}
\label{fig:latticesums-h0.5} 
\end{figure}

The double-tangent construction revealed two different types of phase behavior depending on the value of $H$.  In no case did the configurations resulting from MD or MC yield a ground state. For $0 < H <0.762902$, there exist five distinct structures.  Figure \ref{fig:latticesums-h0.5} illustrates the $u$-$a$ curves for $H=0.5$.  In the figure, the optimal $n$-particle crystals are not visible when they are identical to the triangular, honeycomb, or kagom\'{e} crystals.  As shown in in Fig.\ \ref{fig:latticesums-h0.5}, at highest pressure, a dense triangular lattice (TRI), where neighboring particles lie inside the potential energy barrier of other neighboring particles, is the ground state. At a reduction of pressure to $p=1.84432$, the dense triangular lattice is in coexistence with the honeycomb crystal (HC), which is illustrated with the dashed tangent lines. In Fig.\ \ref{fig:latticesums-h0.5}, the curve for the kagom\'{e} lattice appears possibly to touch the coexistence line due to the finite thickness of the line. However, plotting $p$ against $g$ reveals an intersection between the curves for the TRI and HC structures.  The kagom\'{e} lattice has a higher free energy than both structures in this density range. 

Further reduction of the pressure to 0.75263 yields a ``striped'' (ST) or lane phase coexisting with the honeycomb phase.  This phase is an affinely-stretched triangular lattice. The first few coordination shells contain two, four, and two particles.  The curves representing the lowest-energy crystals with a two- and three-particle basis either the triangular lattice, honeycomb crystal, or kagom\'{e} crystals generally for $a<1.3$ with a corresponding basis.  For larger $a$, these curves appear to come close to the coexistence line between the HC and ST phases or the ST and open TRI phases.  The $p$-$g$ curves reveal that the free energy is always above that of the coexisting phases. These near-ground states are generally crystals that nearly mimic the coexistence. For example, the optimal three-particle crystal between the HC and ST phases is an alternating sequence of lines with honeycomb-like coordination and striped coordination. These are suboptimal structures because the specific neighbor lengths do not match exactly. At sufficiently low pressure, for $H=0.50$ and $p=0.58275$, the ST phase coexists with an ``open'' triangular lattice where the neighbor particles lie on the outside of the potential energy barrier.

Lastly, at vanishing pressure, any configuration in which particles are separated by at least $\rt$ is a ground state since the free-energy vanishes. This occurs for $a\geq 3\rt/2$.  In this region, the ground states behave like hard disks (HD) of radius $\rt$.  There would be a hard-disk crystal and liquid regime with specific area scaled to the hard-disk equation of state, which has been well characterized.\cite{alder1962phase}

\begin{figure}
\includegraphics[width=0.6\textwidth, clip=true]{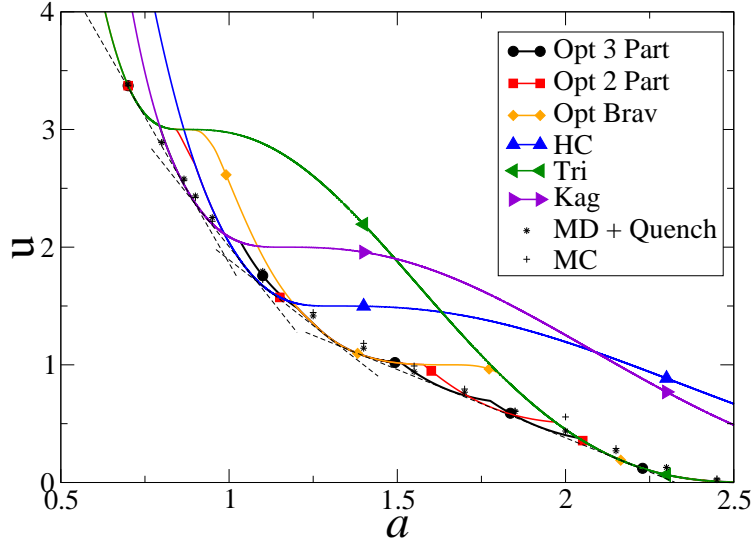}
\caption{(Color online) Potential energy per particle $u$ as a function of specific area $a$ for selected crystals and the optimal 2- and 3-particle crystals for $H=1.00$.  The dense triangular, kagom\'{e}, honeycomb, striped, and open triangular crystals are stable ground states for $H=0.5$ while for $H=1.00$, the kagom\'{e} also emerges as a stable ground state.  The square and anti-kagom\'{e} structures are omitted for clarity. Dashed lines represent the double tangent construction.}
\label{fig:latticesums-h1.0} 
\end{figure}

For $0.762902\geq H \geq 1$, the kagom\'{e} crystal emerges as a ground state structure for a narrow range of stability. This is an especially important result.  In previous work, researchers used an inverse methodology to design pair interactions for targeted ground states.\cite{rechtsman2006designed}  Although they were successful in engineering potential for the honeycomb and square crystals, they were unable to do so for the kagom\'{e}. The area and pressure ranges of stability increase with $H$.  The emergence of the kagom\'{e} crystal as a ground state is shown in the $p$-$a$ curve for $H=1.00$ in Fig.\ \ref{fig:latticesums-h1.0}. As the pressure is reduced, the sequence of stable phases is the dense TRI, KAG, HC, ST, open TRI, and HD phases.

\begin{figure}
\includegraphics[width=0.7\textwidth, clip=true]{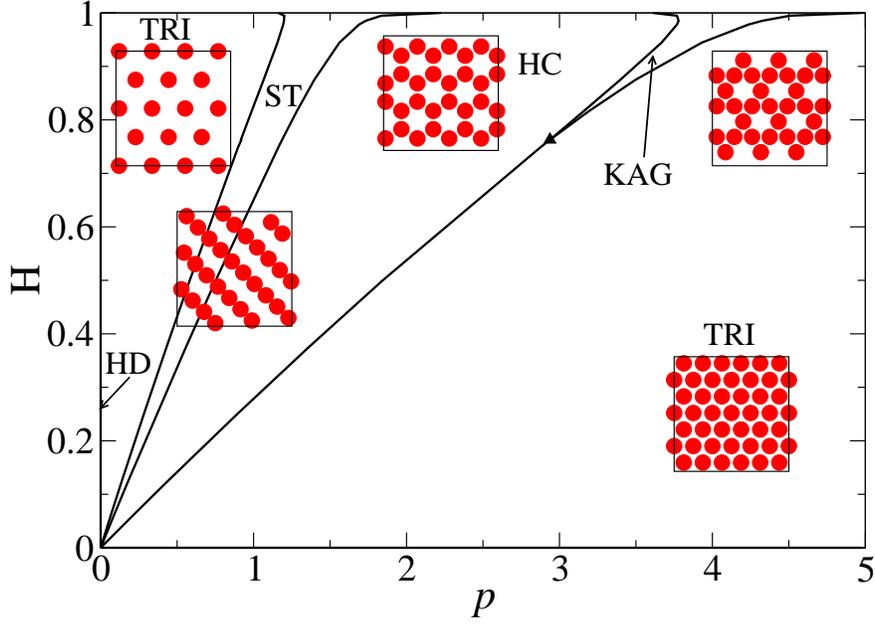}
\caption{(Color online) Phase diagram in the $p$-$H$ plane as calculated by double tangent method.  The hard-disk-like state (HD) occurs for $p=0$ and $H>0$ and the filled triangle on the lower curve represents the triple point.}
\label{fig:phases-ph} 
\end{figure}

\begin{figure}
\includegraphics[width=0.7\textwidth, clip=true]{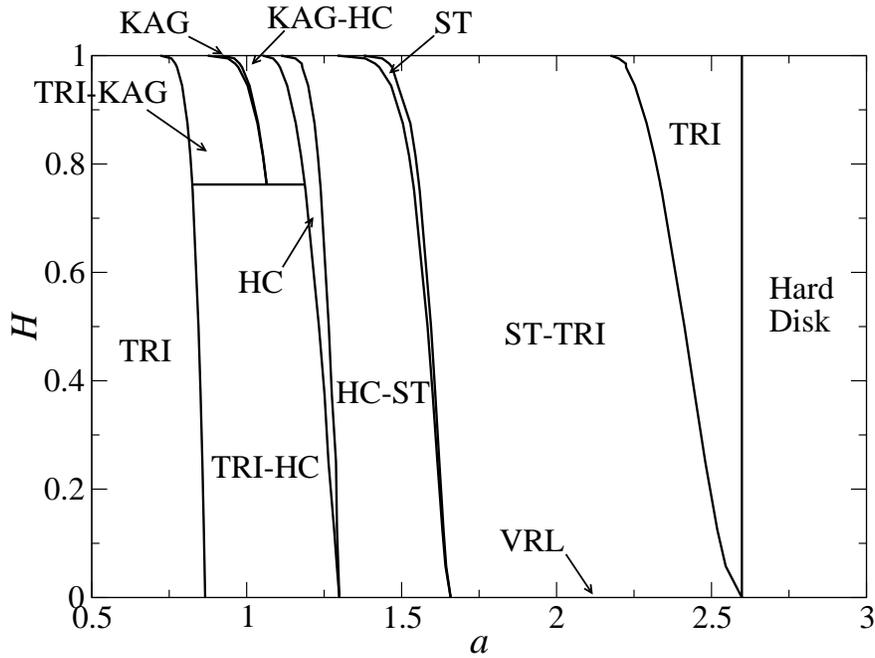}
\caption{Phase diagram in the $a$-$H$ plane as calculated by the double tangent method. }
\label{fig:phases-ah} 
\end{figure}

The full phase diagrams in the $p$-$H$ plane and the $a$-$h$ plane are shown in Figs.\ \ref{fig:phases-ph} and \ref{fig:phases-ah}, respectively.  Figure \ref{fig:phases-ph} shows that the coexistence lines are nearly linear for small $H$.  The pressure range of stability for the open TRI, ST, and KAG phases widens as $H$ increases.   When $H$ is sufficiently large to include the KAG phase, the pressure range of stability for the HC phase narrows.  As $H$ increases, the area range of stability increases for most phases, excluding the dense TRI phase, which is evident in Fig.\ \ref{fig:phases-ah}.

Near $H=1.00$, the slopes of the curves change dramatically.  This is due to the softening of the pair potential function as $H$ approaches unity and the fact that the relative minimum and maximum come together much more rapidly as $H$ approaches unity.  Fig.\ \ref{fig:potential} shows that the $H=1.00$ potential is significantly softer near $r=1$ than other potentials. 

The softening of the potential and the rise of the relative minimum are important features for the inclusion of the KAG phase.  In comparing Fig.\ \ref{fig:latticesums-h0.5} to Fig.\ \ref{fig:latticesums-h1.0}, one can see that the curve for the kagom\'{e} crystal initially begins above the double tangent connecting the TRI and HC phases. As the potential softens and the relative minimum increases, the first minimum in each curve increases proportionally to $H$. However, the softening of the potential bends the left-side of the curves in such a way that the KAG curve lies at lower free energy than that TRI-HC coexistence line.  The combination of the height of the relative minimum and the softening are directly related to the potential energy and pressure of the system, the two components of the $T=0$ free energy.

\begin{figure}
\includegraphics[width=0.7\textwidth, clip=true]{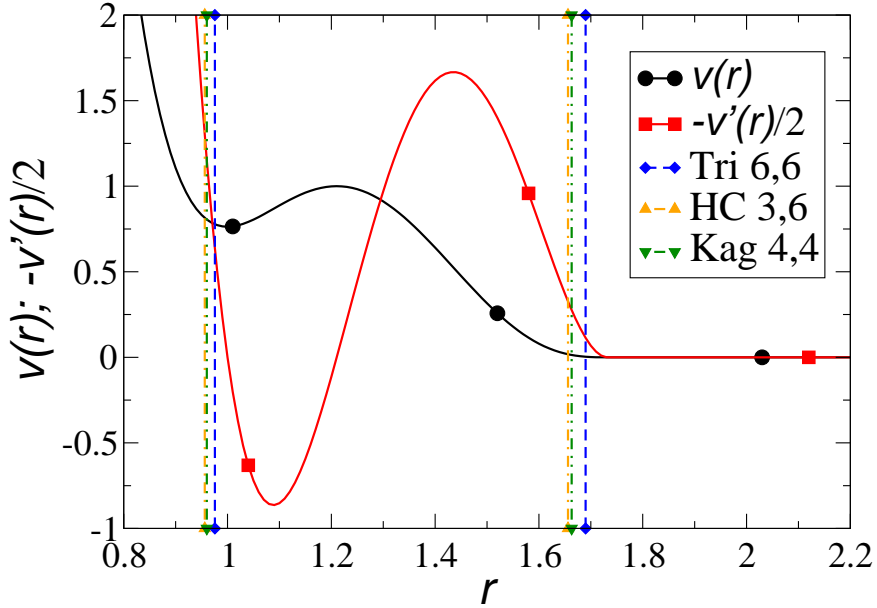}
\caption{(Color online) Comparison of the location and coordination number of the triangular lattice, honeycomb crystal and kagom\'{e} crystal for $a=0.82443$, 1.18745, and 1.06445, respectively, for $H=0.762902$.  At this $H$ and $p=2.938$, the TRI, HC, an KAG phases form a triple point and are in coexistence as ground-state structures. The delicate balance between potential energy and forces allow for coexistence.}
\label{fig:latcompare} 
\end{figure}

The KAG phase emerges at a triple point at $H=0.762909$ and $p=2.938$, where the TRI, HON, and KAG phases form a triple point.  We have examined the subtle interplay between the shape of the pair potential function and its derivative at this coexistence point in Fig.\ \ref{fig:latcompare}. The figure shows $v(r)$ and $-v'(r)/2$ for $H=0.762902$. The plot also shows the location of the nearest and next-nearest neighbors for the crystals. The HC phase has the closest neighbors while the TRI phase has the smallest forces.

The coexistence of these structures requires that a complex system of equations be satisfied. For the coexistence pressure $p^*$ and free energy $g^*$, the system of equations becomes 
\begin{eqnarray}
g^* &=& 3\left[v(d_{t}) + v(d_{t}\rt) - \frac{d_{t}}{2}v'(d_{t}) -\frac{d_{t}\rt}{2}v'(d_{t}\rt)      \right]  \\
p^* &=& -\frac{\sqrt{3}}{d_t}\left[v'(d_t) + \sqrt{3}v'(d_t\sqrt{3}) + \right] \\ 
g^* &=& 3\left[\frac{1}{2}v(d_h) + v(d_h\rt) -\frac{d_h}{4}v'(d_h) -\frac{d_h\rt}{2}v'(d_h\rt) \right]\\ 
p^* &=& -\frac{2}{d_h\rt}\left[ \frac{1}{2}v'(d_h) + \rt v'(d_h\rt)  \right] \\
g^*&=& 2v(d_k) + 2v(d_k\rt) -d_kv'(d_k) - d_k\rt v'(d_k\rt) \\
p^* &=& -\frac{\rt}{2d_k}\left[v'(d_k) +\rt v'(d_k\rt) \right] , 
\end{eqnarray}
where $d_i$ is the nearest-neighbor distance for each structure and is necessarily less than unity for the quintic potential. Evidently, the quintic potential for $H=0.762902$ satisfies these equations. The complexity of the coexistence equations is clear in that each structure weighs certain parts differently.  The design of other potentials that stabilize these structures may utilize this system of equations within an appropriate optimization framework.

\subsection{Stability}

\begin{figure}
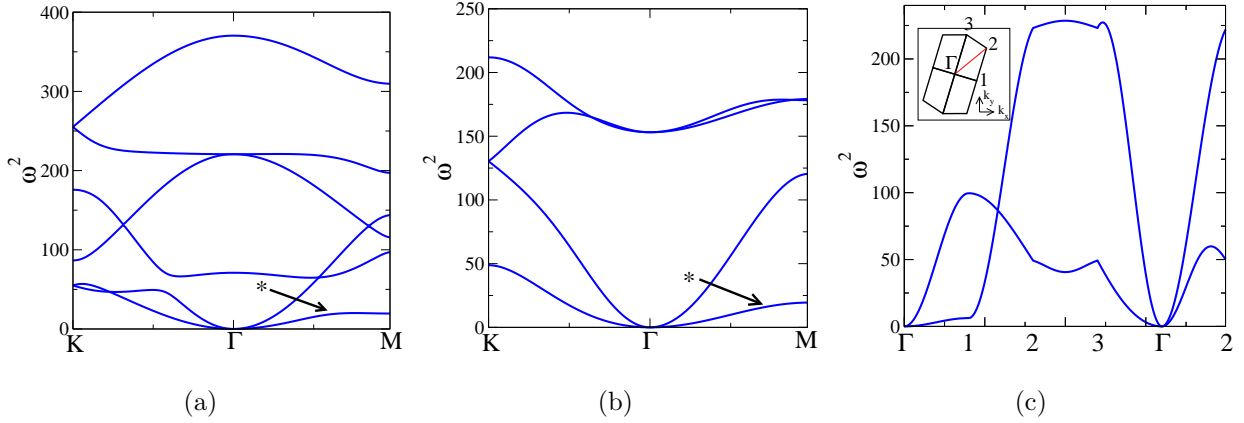

\subfigure[ ]{\label{fig:speckag}
\includegraphics[width=0.32\textwidth, clip=true]{spec_kag_h0.875_a1.035.eps}}
\subfigure[ ]{\label{fig:spechc}
\includegraphics[width=0.32\textwidth, clip=true]{spec_hc_h0.875_a1.2.eps}}
\subfigure[ ]{\label{fig:specst}
\includegraphics[width=0.32\textwidth, clip=true]{spec_st_h0.875_a1.51.eps}}
\caption{(Color online) Phonon spectra for (a) the kagom\'{e} crystal at $a=1.035$ and (b) the honeycomb crystal at $a=1.2$, and (c) the striped phase at $a=1.51$ for wave vectors in the reduced first Brillouin zone for $H=0.875$. The first Brillouin zone for the striped phase is shown in inset in (c). These crystals are mechanically stable at these densities with the quintic potential. The branches containing soft acoustic modes are denoted with a *. }
\label{fig:phonons} 
\end{figure}

The positivity of the squared frequency of all phonons ensures mechanical stability of a crystal. We have examined the phonon spectra for the KAG, HC, and ST phases.  These phases are confirmed to be mechanically stable within the relevant density and $H$ ranges.  Figure \ref{fig:phonons} illustrates the phonon spectra for certain points in the reduced first Brillouin zone for the KAG, HC, and ST structures for $H=0.875$.  THE HC and KAG structures have one particularly soft acoustic branch, labeled in the figures.  Using the ratio $\omega_{max}^2$/$\omega_{min}^2$ at $M$ point a simple metric for the extent of mechanical stability, this ratio stays relatively fixed for the KAG lattice as $H$ increases.  This marks an achievement since previous attempts to stabilize the kagom\'{e} crystal had been unsuccessful.\cite{rechtsman2006designed}  For the HC structure, this ratio is higher for $H=1.00$ than for that which is plotted in Fig.\ \ref{fig:phonons}, indicating that the crystal is more stable as $H$ increases.  In general, the quintic potential and the honeycomb potential\cite{rechtsman2006designed} yield similar ratios.  The striped phase whose lattice vectors for $a=1.51$ and $H=0.875$ are $[1.65029, 0]$ and $[0.27688, 0.91490]$ is also mechanically stable. The phonon spectra from the other ``quadrants'' are identical to the quadrant displayed in Fig.\ \ref{fig:specst} due to the symmetry of the Brillouin zone.

\subsection{Low-Energy Structures}

\begin{figure}
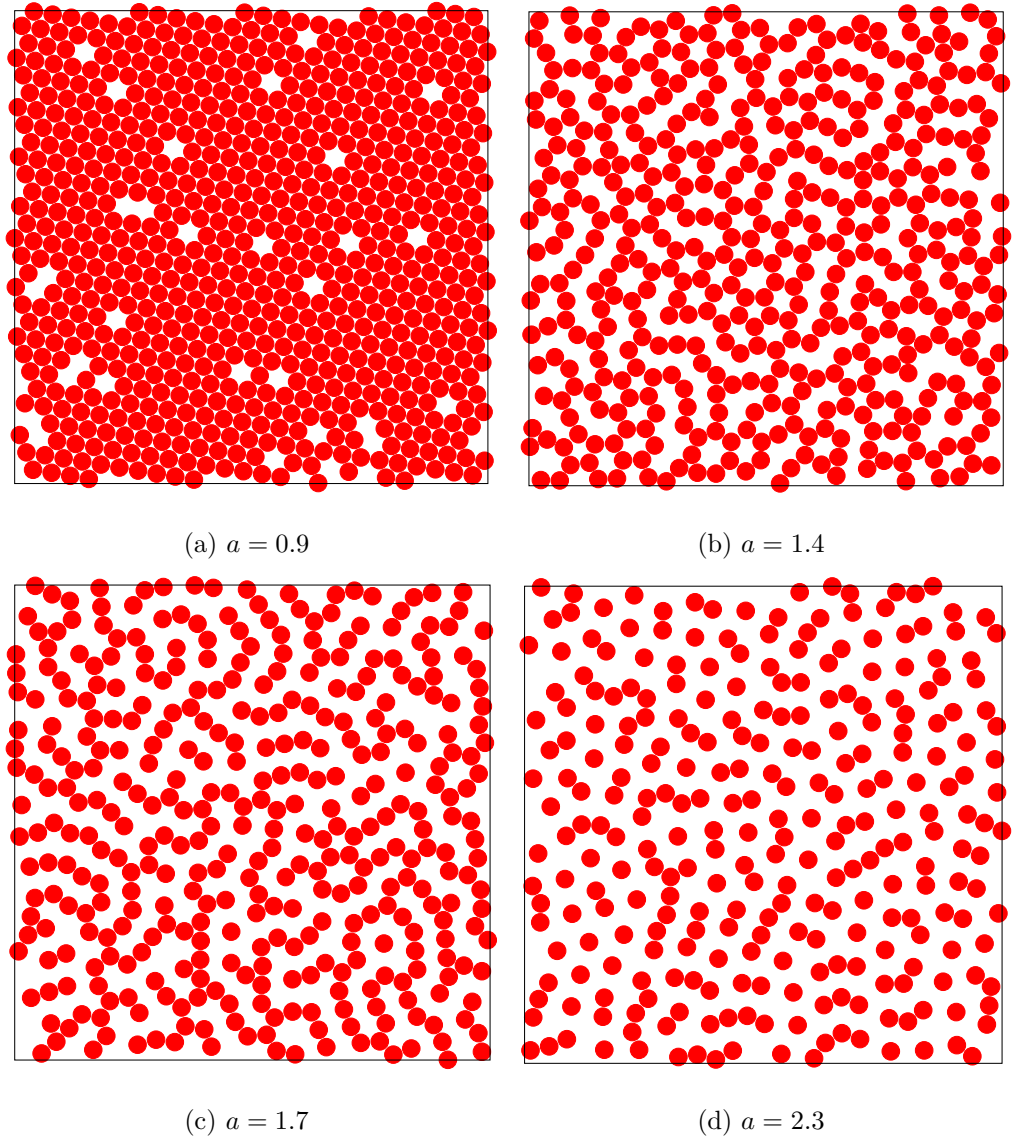

\subfigure[ $\text{ }a = 0.9$]{\label{fig:a0.9}
\includegraphics[width=0.4\textwidth, clip=true]{config_h0.5_a0.9.eps}} 
\subfigure[ $\text{ }a = 1.4$]{\label{fig:a1.4}
\includegraphics[width=0.4\textwidth, clip=true]{config_h0.5_a1.4.eps}} 
\subfigure[ $\text{ }a = 1.7$]{\label{fig:a1.7}
\includegraphics[width=0.4\textwidth, clip=true]{config_h0.5_a1.7.eps}} 
\subfigure[ $\text{ }a = 2.3$]{\label{fig:a2.3}
\includegraphics[width=0.4\textwidth, clip=true]{config_h0.5_a2.3.eps}} 
\caption{(Color online) Metastable configurations obtained via molecular dynamics for $H=0.5$. In an attempt to cool to a liquid to the ground state, the system has difficulty crystallizing and/or phase separating into the appropriate ground-state structures.  As density is decreased, the metastable states go from ordered structures with vacancies to 
a labyrinthine network to a monomeric liquid.}
\label{fig:configs} 
\end{figure}

Although those structure obtained by MD and MC methods were suboptimal, Figures \ref{fig:latticesums-h0.5} and \ref{fig:latticesums-h1.0} show that the free-energy differences between these structures and the ground states are small.  Several structural motifs emerge for this potential and are shown in Fig.\ \ref{fig:configs} for $H=0.5$.  These metastable states may have technological value if the freezing kinetics are sufficiently slow. Preliminary simulations suggest the freezing behavior varies with density.

For example, with $a<1.2$, systems typically freeze into a triangular lattice with vacancies randomly distributed as in Fig.\ \ref{fig:a0.9}.  Although this is not a ground-state structure, it yields a vacancy-riddled lattice.  The vacancies appear to have no particular order. The freezing transition, the temperature at which the system changes from a high-density liquid to an ordered phase, appears to be first order.  The drop in potential energy as the temperature is reduced is sharp in this density range.  

Systems at densities where coexistence between the HC and ST phases are the ground states tend to exhibit a freezing from the liquid phase to a rigid, structured phase. However, the structured phase, which we believe to be metastable, is characterized by rings and strings as in Fig.\ \ref{fig:a1.4}. The six-particle ring is a characteristic of the HC structure and the strings are characteristics of the ST phase.  These metastable states are disordered due to the fluid nature of the strings.  The drop in potential with temperature is not as sharp for these densities compared to those at higher densities.  

Lastly, at lower densities, as with those associated with the ST and open TRI phases, the metastable, low-energy states adopt labyrinthine characteristics, Fig.\ \ref{fig:a1.7}, and eventually colloidal polymers and monomers, \ref{fig:a2.3}. In this density range, the drop in potential energy associated with freezing is weak. The phenomena and characteristics of low-energy states are common to all $H$. However, as $H$ increases so does the propensity to form labyrinthine characteristics.  This is due to the lower energy barrier and the decrease in $r_b$, the location of the energy barrier.  A qualitatively similar, but piecewise linear, potential also yields a labyrinth phase at positive temperature.\cite{haw2010growth} Although the positive temperature behavior has not been fully explored in this paper, we anticipate that the equation of state for this family of potentials will have interesting behavior. Manipulation of the kinetics to achieve such unusual structures represents one path to achieving kinetically stable materials with controlled vacancy concentrations.

\section{Special Cases: $H=0$}
\label{sec:cases}

The $H=0$ pair potential has interactions that vanish at a pair distance of unity and beyond $\rt$.  At positive pressure, the ground state is the dense triangular lattice. However, at zero pressure and $a\geq \rt/2$, ground-state configurations will have a vanishing potential energy and pressure (enthalpy vanishes). Thus, the type of available ground states is dependent on the area. 

A number of interesting structures arise as ground states.  For $a>\rt/2$, dilutions of the triangular lattice with unit neighbor spacing are ground states, including the honeycomb and kagom\'{e} crystals and other lattice gases. For $a = \sqrt{11}/2$, the striped phase, a Bravais lattice with lattice vectors of $[1, 0]$ and $[1/2, \sqrt{11}/2]$, becomes available as a ground state.  Each particle has two neighbors at unit separation and four neighbors at separation of $\rt$.  For $a > \sqrt{11}/2$, dilutions of this lattice are also ground states. In addition, any small expansion in the direction normal to the stripes, will allow each stripe to gain some fluidity.  Complex liquids of strings have found as ground states using molecular dynamics for at least $a\geq 2.15$. However, the geometric problem is highly nontrivial since for several runs with $a>2.15$, ground states could not always be obtained. For $a=3\rt/2$, an open triangular lattice with neighbor spacing of $\rt$ is a ground state structure.  

The question remains as to which type of configurations are entropically (thermodynamically) favorable, or rather which type of system makes up the largest fraction of configuration space.  We first make the distinction between discrete ({\it e.g.} lattice gas) and continuous entropy ({\it e.g.} hard-disk crystal). In the classical case, continuous entropy is uncountable while discrete entropy is not. Making the analogy to hard disks and hard spheres, we believe that the highest entropy phases for a specified area are those that have the availability for continuous entropy.  Configurations with the highest-dimensional configuration space dominate the $T=0$ entropy. We estimate that, for a certain $a$, configurations with the fewest constrained degrees of freedom will have the highest dimensional configuration space, and therefore the highest entropy. For $a\leq \rt/2$, all degrees of freedom per particle are constrained.   For $a\geq3\rt/2$, no degrees of freedom per particle are constrained.

In order to gain continuous entropy, the system must have the ability to expand the $\rt$ ``bonds,'' which can be either first-neighbor or second-neighbor connections between particles.  Any slight expansion of the dense triangular lattice is not a ground state because each neighbor must be constrained to unit separation. However, an expansion of the open triangular lattice is still a ground state because the $\rt$ bonds are not constrained from above. The problem can be cast as a tiling problem of four triangular tiles.  The possible triangular tiles are those with side lengths of unity or $\rt$.  These are depicted in Table \ref{tab:tiles} along with their shorthand notations. 

Because there are no constraints beyond distances of $\rt$, the edges with lengths of $\rt$ can be considered ``elastic.'' For simplicity, we first consider the tiling problem where these edge lengths are fixed. The rules of the tiling problem are as follows:
\begin{itemize}
 \item The plane must be tiled with no gaps.
 \item For pairs of tiles that share an edge, the edges must be of the same length so that the vertices also match.
 \item The (1,1,r) and (1,r,r) tiles cannot share a $\rt$ edge since this violates a second neighbor constraint ({\it i.e.} two vertices are separated by a length between unity and $\rt$).
\end{itemize}

\begin{table}
\caption{\label{tab:tiles} Available triangular tiles for the $H=0$ tiling problem.}
\begin{ruledtabular}
\begin{tabular}{l|cll}
Name        & Shape  & Angles ($^\circ$) & Area \\ \hline
(1,1,1)       & \includegraphics[scale =0.4,clip=true]{111.eps} & 60, 60, 60 & $\rt/4$ \\ \hline
(r,r,r)       & \includegraphics[scale =0.4,clip=true]{rrr.eps} & 60, 60, 60 & $3\rt/4$ \\ \hline
(1,1,r)       & \includegraphics[scale =0.4,clip=true]{11r.eps} & 30, 30, 120 & $\rt/4$ \\ \hline
(1,r,r)       & \includegraphics[scale =0.4,clip=true]{1rr.eps} & 33.56, 73.22, 73.22 & $\sqrt{11}/4$ \\
\end{tabular}
\end{ruledtabular}
\end{table}

\begin{figure}
\subfigure[]{\label{fig:denseopen}
\includegraphics[width=0.3\textwidth, clip=true]{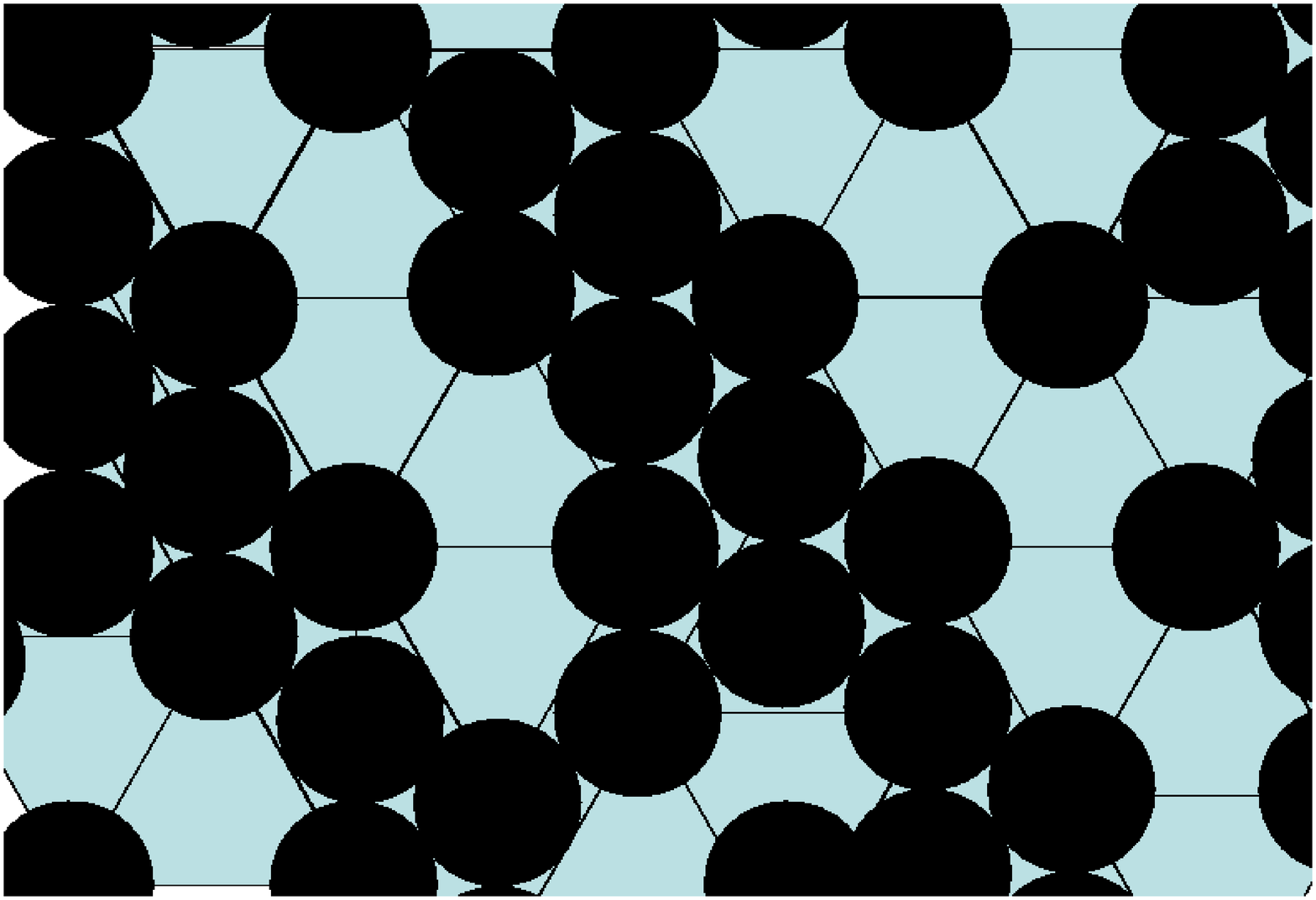}}
\subfigure[]{\label{fig:stripe}
\includegraphics[width=0.3\textwidth, clip=true]{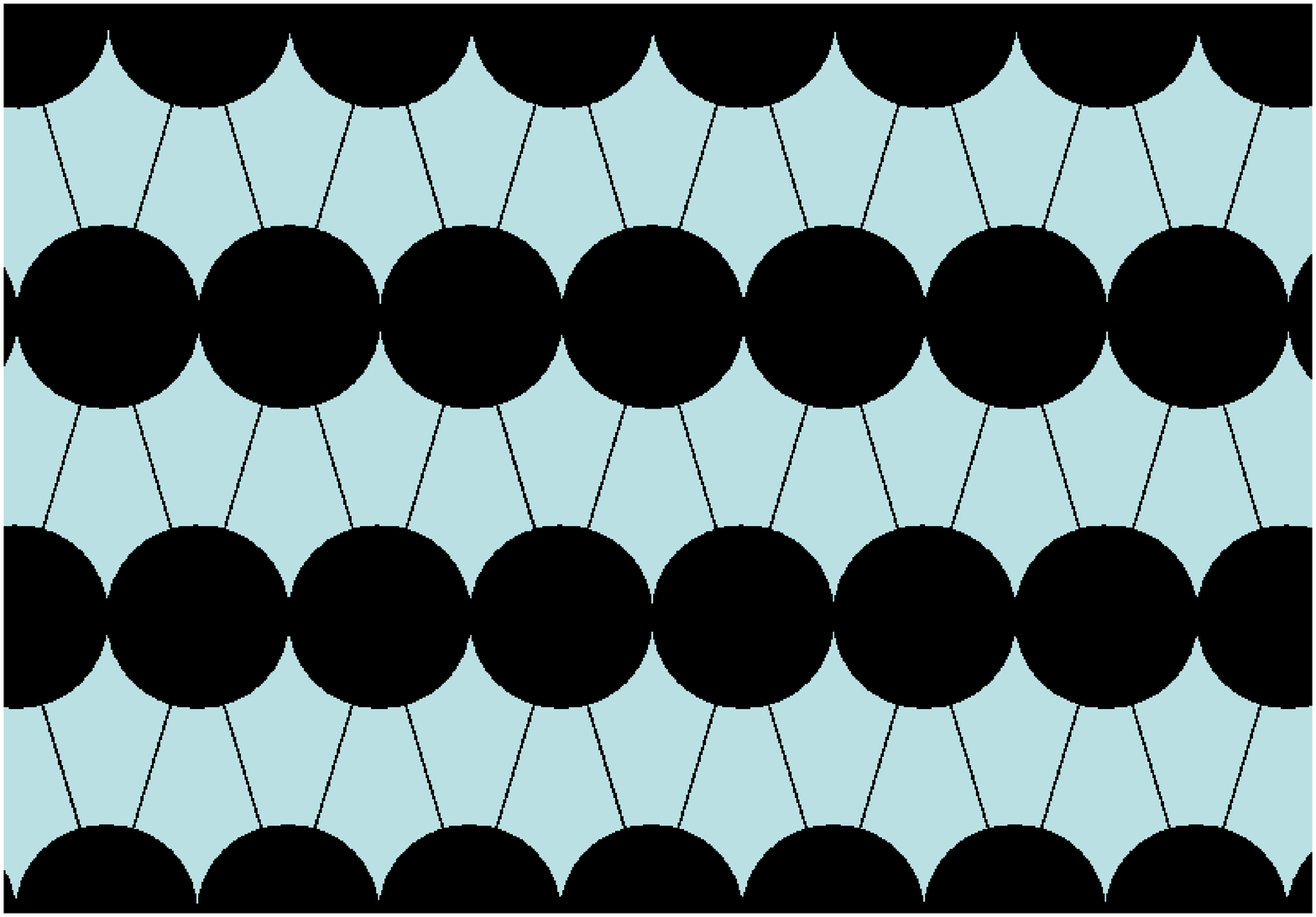}}
\subfigure[]{\label{fig:zigzag}
\includegraphics[width=0.3\textwidth, clip=true]{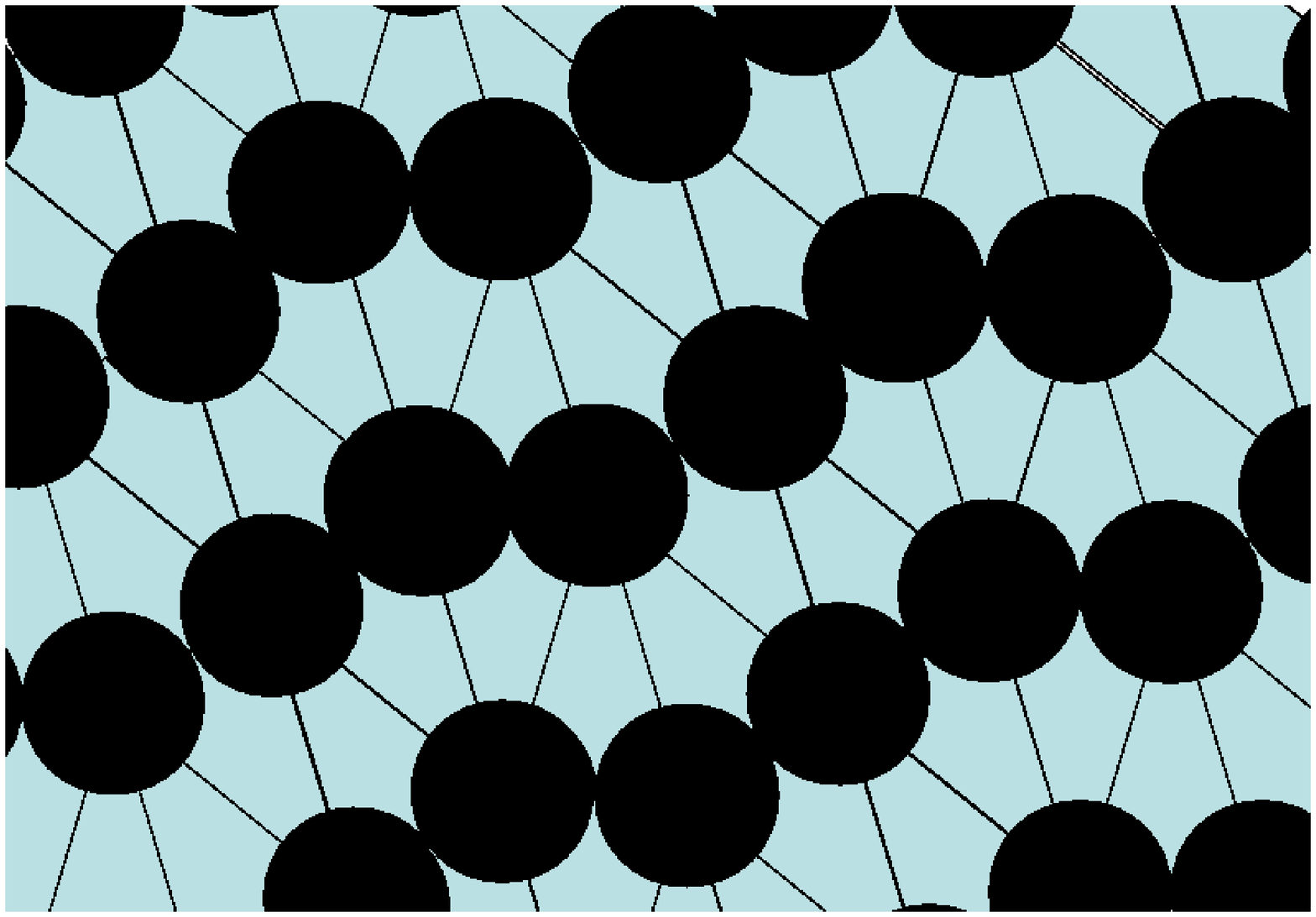}}
\subfigure[]{\label{fig:densestripe}
\includegraphics[width=0.3\textwidth, clip=true]{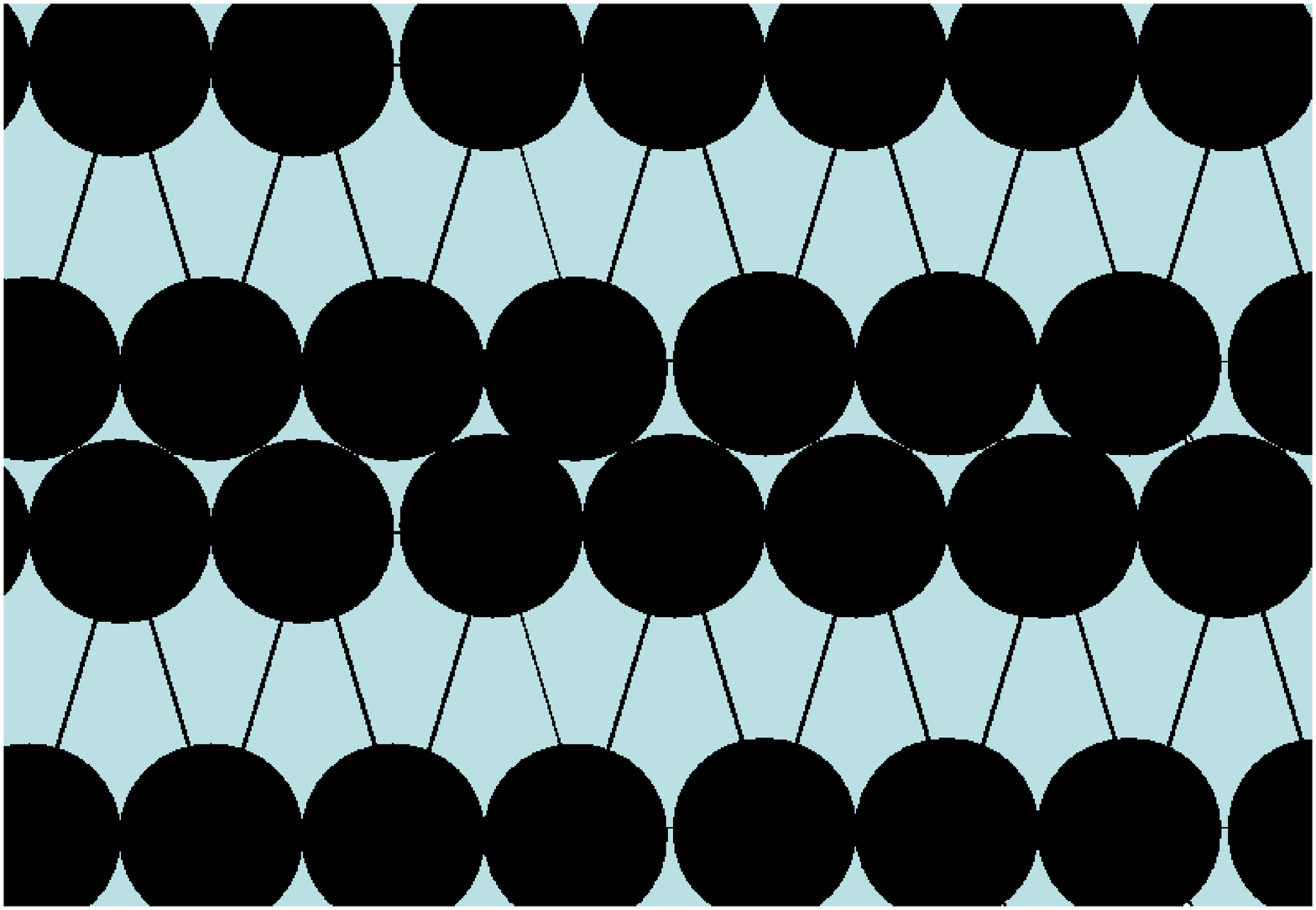}}
\subfigure[]{\label{fig:openstripe}
\includegraphics[width=0.3\textwidth, clip=true]{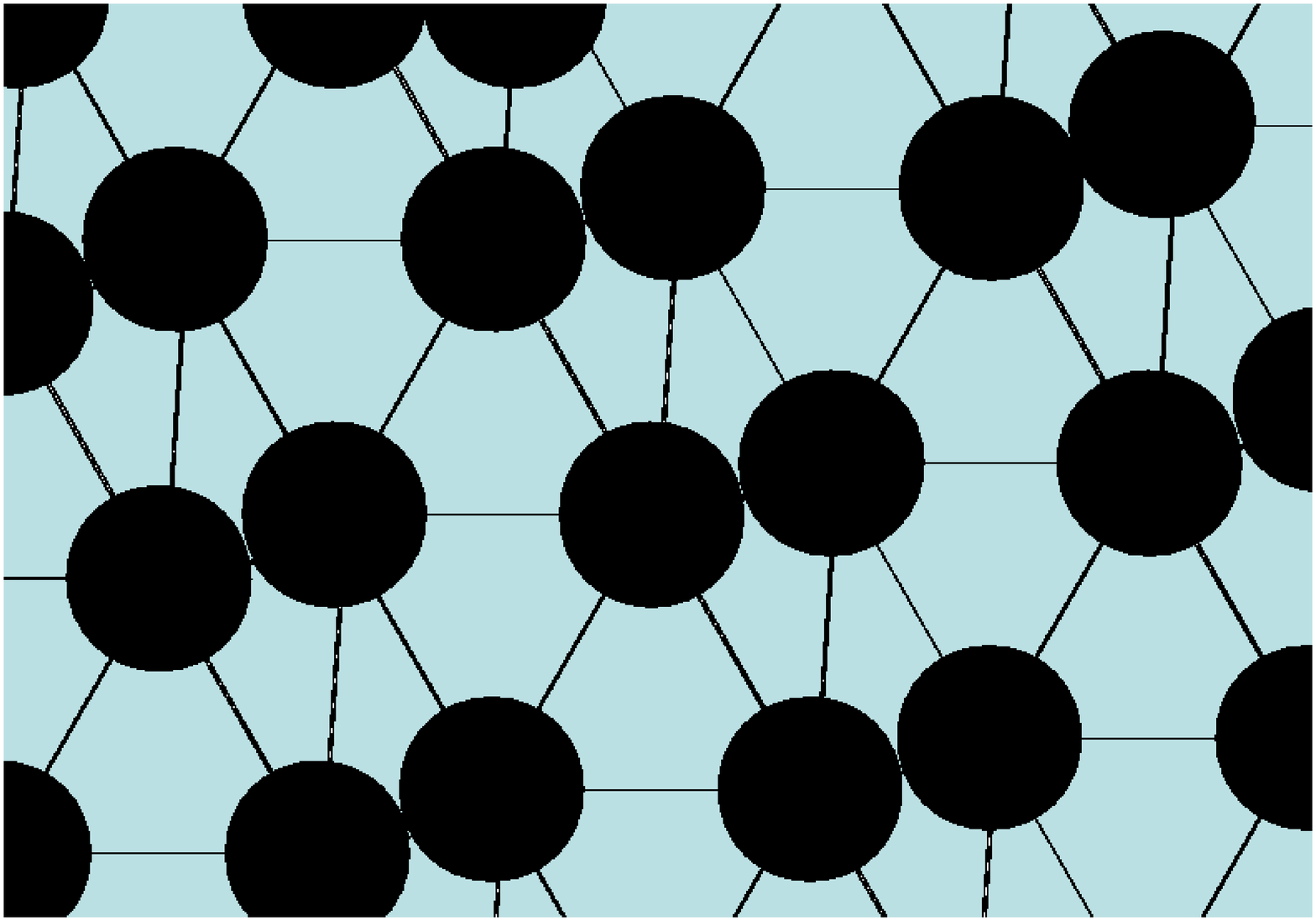}}
\caption{(Color online) (a) Tiling excluding (1,r,r) tiles, (b) striped phase using only (1,r,r) tiles, (c) zig-zag phase using only (1,r,r) tiles (d) coexistence between dense triangular lattice and the striped phase, and (e) coexistence between the open triangular lattice and the striped phase.}
\label{fig:tilings} 
\end{figure}

\begin{figure}
\includegraphics[width=0.7\textwidth, clip=true]{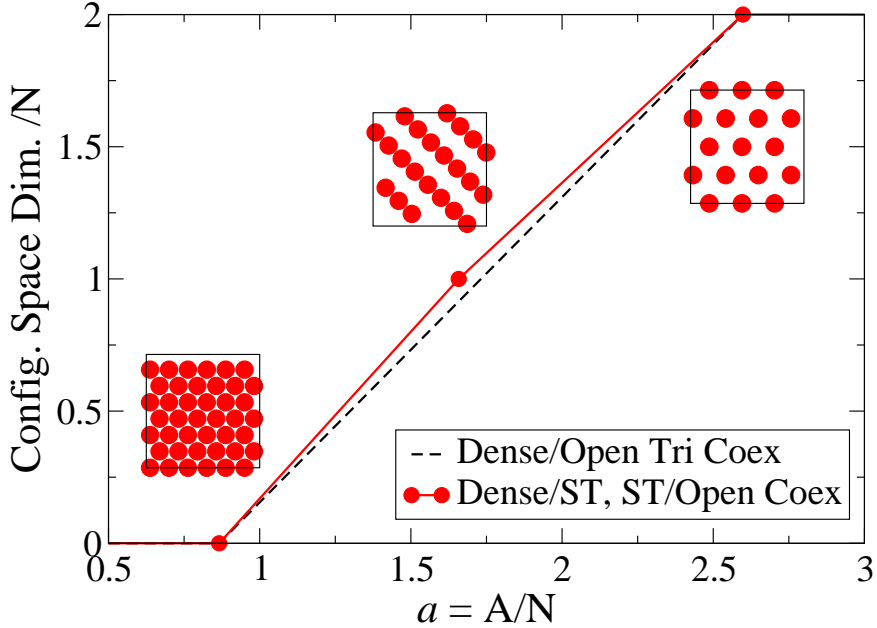}
\caption{(Color online) Unconstrained degrees of freedom per particle for candidate ground-state structures for $H=0$ potential.  Configurations with the maximum number of unconstrained degrees of freedom are presumed to be the ground state (red circles).  }
\label{fig:dimension} 
\end{figure}

The internal angle of the (1,r,r) tiles are such that they cannot integrate well with the other tiles. Therefore, in any tiling that includes (1,r,r) tiles, these tiles must ``phase separate'' from the others.  We consider two types of tilings - those with (1,r,r) tiles and those without. First, we consider those without (1,r,r) tiles. An example of such a tiling is shown in Fig.\ \ref{fig:denseopen} where each vertex is decorated with a particle.  These tilings are simply dilutions of the dense triangular lattice, or coexistence between a dense and open triangular lattice. The (1,1,r) tiles have two roles. They can act as intermediaries between domains of the dense triangular lattice and the open triangular lattice, and they can dimerize to form a rhombus making a domain of the dense triangular lattice.  This coexistence between the dense and open triangular lattices is represented as a double-tangent line connecting the fully constrained dense triangular lattice to the unconstrained open triangular lattice as shown in Fig.\ \ref{fig:dimension}. 

Next we consider those tilings that include (1,r,r) tiles.  Two periodic tilings with a specific area of $a= \sqrt{11}/2 \approx 1.658312$ exist and are shown in Figs.\ \ref{fig:stripe} and \ref{fig:zigzag}.  We deem these the stripe and ``zig-zag'' tilings respectively. In these phases, any small expansion perpendicular to the stripe or zig-zag will allow the stripes to have fluidity.  On average, these configurations will constrain one degree of freedom per particle, as plotted in Fig.\ \ref{fig:dimension}.  

For any tiling consisting of all tiles, the (1,r,r) tiles must segregate so that the the other tiles can fully tile the plane.  Dimerizing the (1,r,r) tiles along the $\rt$ edges creates a wide stripe.  Building off the wide stripe requires (1,1,1) tiles or (1,1,r) tiles and forms a local coexistence with the dense triangular lattice as shown in Fig.\ \ref{fig:densestripe}.  Dimerizing the (1,r,r) tiles along the unit length edge creates a narrow stripe. The exposed edges have length $\rt$, requiring (r,r,r) tiles to be directly adjacent to the stripe. This forms a coexistence between the the striped phase and the open triangular lattice. The double tangent construction of these coexistences is displayed as the solid red line in Fig.\ \ref{fig:dimension}.  These coexistences between the striped phase and the dense triangular lattice and the striped phase and the open triangular lattice maximize the number of unconstrained degrees of freedom and likely maximize the entropy of the ground state.  In addition, these system can take on discrete entropy by way of stacking variants of striped phase and the lattice phase.  Using ``S'' and ``T'' to denote a stripe and a triangular lattice layer, a STTTTSSS system is degenerate with a STSTSTST system.

The zig-zag phase can only form local coexistence with the open triangular lattice. To do so, the (1,r,r) tiles must form a trimer which exposes the $\rt$ edges on either side of the trimer.  Since they are in trimers, their stacking entropy would be less that the stacking entropy available to the coexistence between the stripe and open triangular lattice. Therefore, we believe the stripe phase has a higher entropy than the zig-zag phase.

\section{Discussion and Concluding Remarks}
\label{sec:discussion}

In this paper, we have developed the ground-state phase diagram of a new potential that gives rise to a number of novel phases that include low-coordinated crystal structures.  
Given the unusual nature of the ground state, we expect that the equation of state and full phase diagram for these systems to exhibit other unusual behaviors at positive temperature.  For example, in the global phase diagram of the the Lennard-Jones-Gauss, or honeycomb potential, there was no gas-liquid coexistence in the low-temperature, low-density part of the phase diagram, nor was there a liquid-liquid phase coexistence\cite{hynninen2006global} We expect most of the solid-solid transitions that we find at zero temperature will remain at small nonzero temperature. Our preliminary calculations suggest that strings, or polymers, may arise in equilibrium at low densities. 

In addition, we have interest in the mobility of vacancies, particularly in the cases where vacancy concentrations are dilute, ({\it e.g.} $H=0$ and $a\approx \rt/2$), due to the possible relation to supersolid behavior.\cite{andreev1969soviet,kim2004probable}  For $a$ just above $\rt/2$ and $H=0$, we expect there to be a small number of vacancies in the system at low-temperature.  We have estimated the potential energy required for a particle to ``hop'' from a lattice site to a vacant site. By initializing a system with one vacancy
with a ``hopping'' particle along the transition to the vacant site, in an otherwise undistorted lattice, we minimized $F({\bf r}^N)=|\nabla {\bf u}({\bf r}^N)|^2$, where ${\bf r}^N$ represents particle coordinates, using conjugate gradient minimization. The resulting configuration represents a saddle point in the energy landscape. The difference in potential energy of the saddle point and the ground state is the energy barrier required for the particle and vacancy to swap positions.  This barrier sets an activation energy for classical thermal motion. For the quantum case, it would also enter in the tunneling rate.  For $H=0$ potential, there is a single saddle point in the vicinity of the numerous initial conditions in the energy landscape with a total energy barrier height of 5.569015. The diffusing particle is midway between the origin and destination sites while the bracing particles are displaced off the lattice line to accommodate the jump.  Using molecular dynamics, we expect to relate this saddle point energy to a vacancy diffusion coefficient.

The quintic potential can further be generalized by varying the distance at which the function is truncated. We set this cutoff distance to be $\rt$. However, allowing this cutoff distance to vary introduces a larger class of potentials. A systematic study of the cutoff radius on the robustness of the ground states is necessary for experimental realization. The hard-core plus square shoulder potential has ground states that vary significantly depending on the relative lengthscale of the hard-core distance and the square-shoulder distance\cite{fornleitner2008lane,fornleitner2010pattern} It is expected that the ground states of the generalized quintic potential would be sensitive to the location of the minimum and the cutoff distance, though it is currently unknown how sensitive the ground-state phase diagram is to these parameters.  Understanding this sensitivity is important to experimentalists who want a simple, robust potential.  Developing an optimization procedure to make self-assembly more robust would be particularly useful. In addition, an extension to three-dimensions may provide additional fundamental insight.

As mentioned earlier, inverse optimization techniques have been effective in developing potentials for targeted material properties. We intend to develop a general and broad inverse optimization technique to target specific vacancy arrangements by accounting for and/or manipulating long-ranged vacancy-vacancy interactions. For example, one might develop an objective function whose variables include the strength, sign (attractive/repulsive), and angular dependence of vacancy-vacancy interactions.  Alternatively, one might consider a two-component system system of a heavy particle and a light particle and apply an inverse optimization technique to this system. Using a broad family of potentials, one could then optimize over the available parameters to achieve a dilute concentration of effectively repulsive vacancies.

\section{Acknowledgments}
We thank Lawrence Cheuk for initial work on a related model. S.T. thanks the Institute for Advanced Study for its hospitality during his stay
there. This work was supported by the Office of Basic Energy Sciences, U.S.
Department of Energy, Grant DE-FG02-04-ER46108..

\bibliography{VacancyGS}
\bibliographystyle{rsc}

\end{document}